\documentclass{article}
\usepackage{frascatiphys}
\usepackage{graphicx}

\begin{document}
\title{ 
SPECTRA OF ASTROPHYSICAL PARTICLES AT \\ 
ULTRA HIGH ENERGY AND THE AUGER DATA}
\author{
Roberto Aloisio \\
{\em INAF - Osservatorio Astrofisico di Arcetri, Firenze, Italy} \\ 
{\em Gran Sasso Science Institute (INFN), L'Aquila, Italy} \\
}
\maketitle
\baselineskip=11.6pt
\begin{abstract}
We use a kinetic-equation approach to describe the propagation of ultra high energy cosmic ray protons and nuclei comparing theoretical results with the observations of the Pierre Auger Observatory. 

\end{abstract}
\baselineskip=14pt
\section{Introduction}

After more than hundred years from the discovery of Cosmic Rays (CR), performed through the first balloon flight of Victor F. Hess in 1912, we have unveiled and understood most of the fundamental aspects of  this fascinating phenomenon. In the energy range that spans from few GeV/n up to $10^{3}$ TeV/n a self consistent scenario that accommodates CR composition, propagation and sources was developed in the last 30 years, the so-called standard model of galactic CR (for a review see \cite{Ptuskin:2012vs} and references therein). At the highest energies, in the regime of Ultra High Energy Cosmic Rays (UHECR) with $E>10^{17}$ eV, the situation becomes much more unclear. The origin of UHECR, with observed energies up to $10^{20}$ eV, has been challenging our theoretical understanding since long time and a clear solution of the problem is still lacking.  

At energies in the range $10^{17}\div 10^{19}$  eV the propagation of UHE particles is extended over cosmological distances with a typical path length of the order of Gpc. Therefore the particle's energy is affected by the cosmological expansion of the universe that results in an adiabatic process of energy loss. Together with cosmology the propagation of UHECR is affected by the interaction with astrophysical radiation fields: the Cosmic Microwave Background (CMB) and the background field constituted by infra-red, visible and ultra-violet radiation that we will call collectively: Extragalactic Background Light (EBL). The first background is the well known cosmic radiation fossil of the big bang, with a black body spectrum degraded nowadays to a temperature of $2.7$ $^\circ$K. The EBL radiation field is produced by astrophysical objects at present and past cosmological epochs and subsequently modified by red-shift and dilution due to the expansion of the Universe. 

As stated above the propagation of UHECR extends over cosmological distances, thus the cosmological evolution of the backgrounds has a non negligible role in the propagation of UHECR. While the cosmological evolution of the CMB is analytically known, the evolution with red-shift of the EBL field should be inferred from observations at different red-shifts through specific models \cite{EBLmodels}. These models are in good agreement in the low red-shifts regime ($z<4$), which is the most important in the physics of UHECR, while show significant differences in the high red-shift regime ($z>4$). These differences have a small impact on the expected UHECR flux and composition they affect only the production of cosmogenic neutrinos (see \cite{Allard:2011aa} and references therein). Being this issue outside the aims of the present paper we will not enter this discussion here using the recipe of Stecker \cite{EBLmodels} to model the EBL radiation field and its cosmological evolution. 

Let us now briefly recall the dominant interactions channels of UHECR with the background radiation, with particular emphasis on their imprints on the expected flux. We will restrict the discussion to charged particles, being the possibility of photons as UHECR very unlikely as follows from experimental observations \cite{GammaLimits}. 

The propagation of UHE nucleons\footnote{Hereafter we will refer only to protons because, as discussed in \cite{Allard:2011aa,Aloisio:2008pp,Aloisio:2010he}, the decay time of neutrons is much shorter than all the other time scales involved in the propagation of UHE particles.} is affected only by the interaction with the CMB radiation field \cite{Aloisio:2008pp,Aloisio:2010he}. There are two spectral signatures that can be firmly related to the propagation of protons through this background: pair-production dip \cite{Berezinsky:2002nc}, which is a rather faint feature caused by the pair production process: $p+\gamma_{CMB} \to e^{+} + e^{-} + p$, and a sharp steepening of the spectrum caused by the pion photo-production: $p+\gamma_{CMB} \to \pi + p$ called Greisen-Zatsepin-Kuzmin (GZK) cut-off \cite{GZK}. The GZK cutoff position is roughly defined by the energy where the pair-production and the photo-pion production energy loss become equal, namely at $E_{GZK} \simeq 50$ EeV \cite{Berezinsky:1988wi}.

The propagation of UHE nuclei, apart from CMB, is affected also by the EBL. The interaction processes that condition the propagation of UHE nuclei are pair production, that involves only the CMB background \cite{Aloisio:2010he,Aloisio:2012aa,Aloisio:2012ba}, and photo-disintegration. The latter is the process in which a nucleus of atomic mass number $A$ because of the interaction with CMB and EBL looses one or more nucleons $A+\gamma_{CMB,EBL} \to (A-nN) + nN$, being $n$ the number of nucleons lost by the nucleus \cite{Aloisio:2008pp,Aloisio:2010he,Aloisio:2012aa,Aloisio:2012ba}.The photo-disintegration of nuclei, together with the pair production process, produces a steepening in the observed spectrum. The exact position of the flux suppression depends only on the nuclei species, being a consequence of the interaction with the CMB field thus free from the uncertainties connected with the EBL \cite{Aloisio:2008pp,Aloisio:2010he,Aloisio:2012aa,Aloisio:2012ba}. 

From the experimental point of view the observations of UHECR are far from being clear with different experiments claiming contradictory results. The HiRes and, nowadays, the Telescope Array (TA) experiments show a proton dominated composition till the highest energies with a clear observation of the proton pair-production dip and GZK cut-off \cite{TAChem,HiResChem}. Chemical composition observed by HiRes and TA is coherent with such picture showing a pure proton dominated spectrum starting from energies $E\simeq 10^{18}$ eV till the highest energies. The experimental picture changes taking into account the Auger observations. The spectrum observed by Auger\footnote{Here we consider the Auger data published in 2011, the new results published in 2013 do not change the picture presented here \cite{Aloisio:2013hya}.} shows a behavior not clearly understood in terms of the proton pair-production dip and GZK cut-off \cite{AugerFlux}. This spectral behavior could be a signal of a substantial nuclei pollution in the spectrum, which is confirmed by the Auger observations on chemical composition that show a progressively heavy composition toward the highest energies, this tendency starts already at $E> 4\times 10^{18}$ eV \cite{AugerChem}. 

In the present paper we will restrict our analysis only to the Auger data discussing the assumptions on the sources of UHECR that enable a good description of the flux \cite{AugerFlux} and chemical composition \cite{AugerChem} observed by Auger. This analysis is performed without a formal fitting procedure, based on some likelihood method, that would be rather time consuming given the analytic computation scheme used. The main goal of this paper, based on the analysis presented in \cite{Aloisio:2013hya}, is to give an overall picture, inferring general rules about the possible sources of UHECR.

The paper is organized as follows: in the next session \ref{sec:Auger} we will focus on the Auger data discussing the source models that better reproduce observations in terms of both flux and chemical composition, we will conclude in section \ref{sec:conclude} discussing the main outcomes of our study. 

\section{Auger Observations} 
\label{sec:Auger}

Following the discussion recently presented in \cite{Aloisio:2013hya} we will use the theoretical framework based on the kinetic approach for the propagation of UHE particles, that was introduced in \cite{Berezinsky:1988wi} for protons and in \cite{Aloisio:2008pp,Aloisio:2010he} for nuclei. As in \cite{Aloisio:2013hya}, we will assume that the spectrum of the accelerated particles at the sources has a power law behavior in energy and the sources are homogeneously distributed in the Universe with no cosmological evolution. Through this simplified theoretical approach we will compare Auger data on flux and chemical composition with theoretical predictions finding interesting general consequences on source models. 

\begin{figure}[htb]
    \begin{center}
        \includegraphics[scale=0.46]{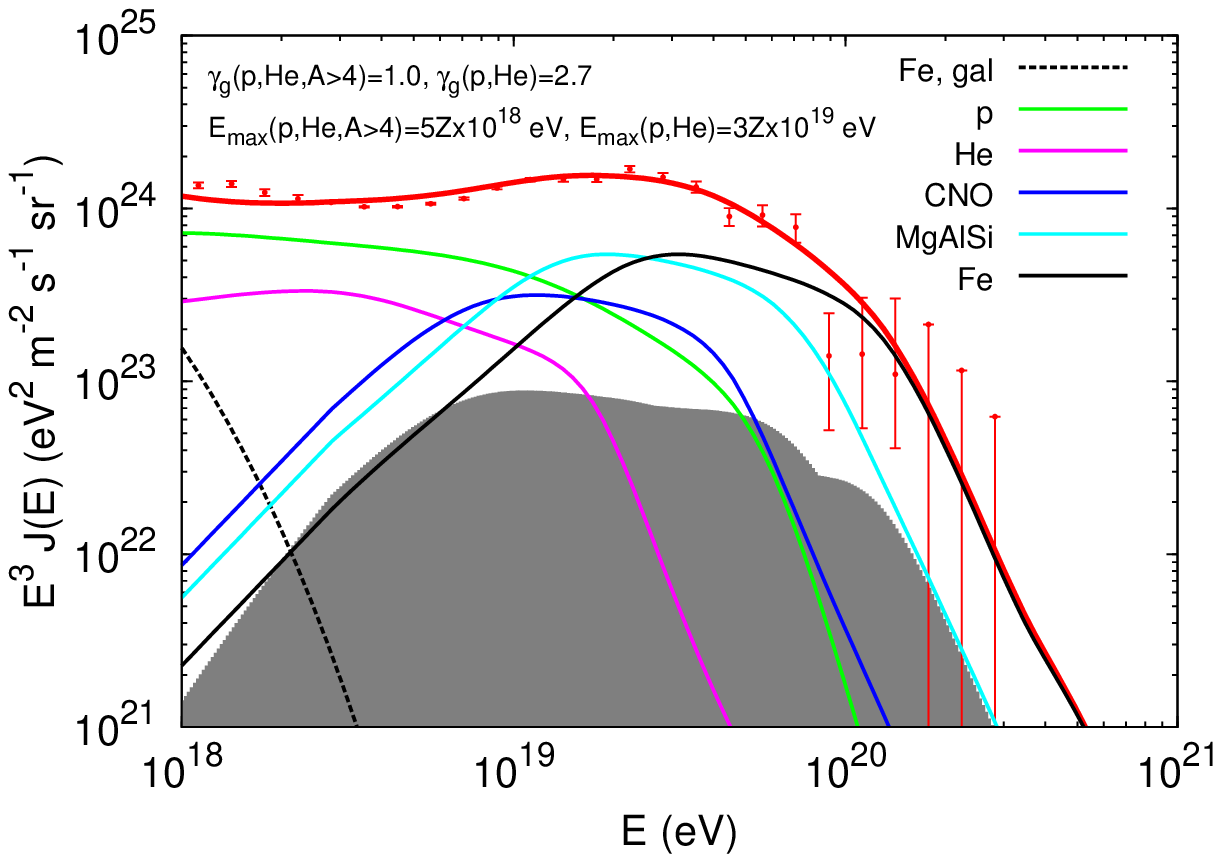}
        \includegraphics[scale=0.46]{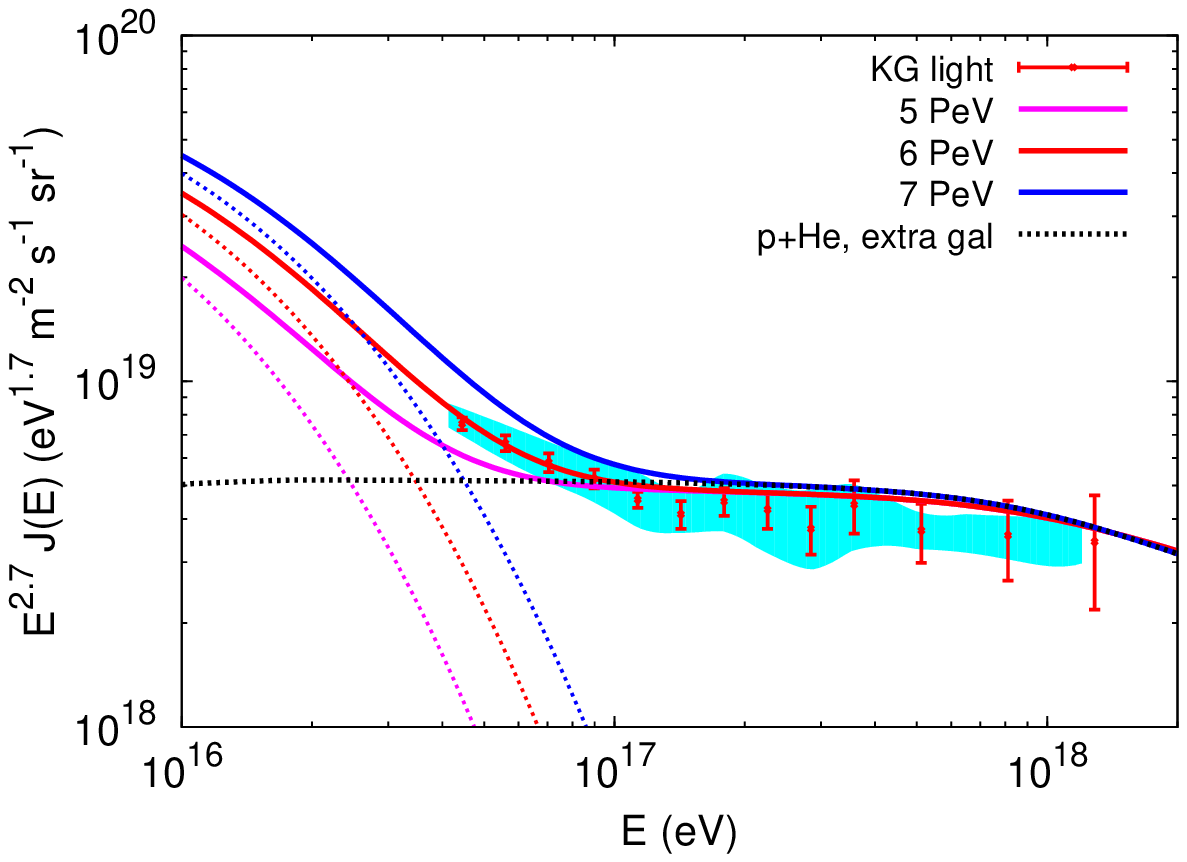}
        \caption{\it [Left Panel] Flux of UHECR obtained with two classes of sources as described in the text in comparison with the flux observed by Auger \cite{AugerFlux}. [Right Panel] Low energy tail of the light UHECR component in comparison with the KASCADE-Grande data on the light component in the high energy tail of galactic cosmic rays \cite{KG}.}
\label{flux}
    \end{center}
\end{figure}

The most commonly used shower observables to study the composition of UHECR are the mean value of the depth of shower maximum $\langle X_{max} \rangle$ and its dispersion $\sigma(X_{max})$. As was first discussed in \cite{Aloisio:2007rc}, the combined analysis of $\langle X_{max} \rangle$ and $\sigma(X_{max})$ is more sensitive to chemical composition and provides less model dependent results. However, inferring chemical composition of UHECR from these observables is subject to some level of uncertainty because their conversion to mass relies on shower simulations codes which depend on the assumptions on the hadronic interaction models. These models, while give the same fit to low energy accelerator data, provide different results of the high energy extrapolations needed in UHECR physics (for a review see \cite{Engel:2011zzb} and references therein).  

\begin{figure}[htb]
    \begin{center}
        \includegraphics[scale=0.46]{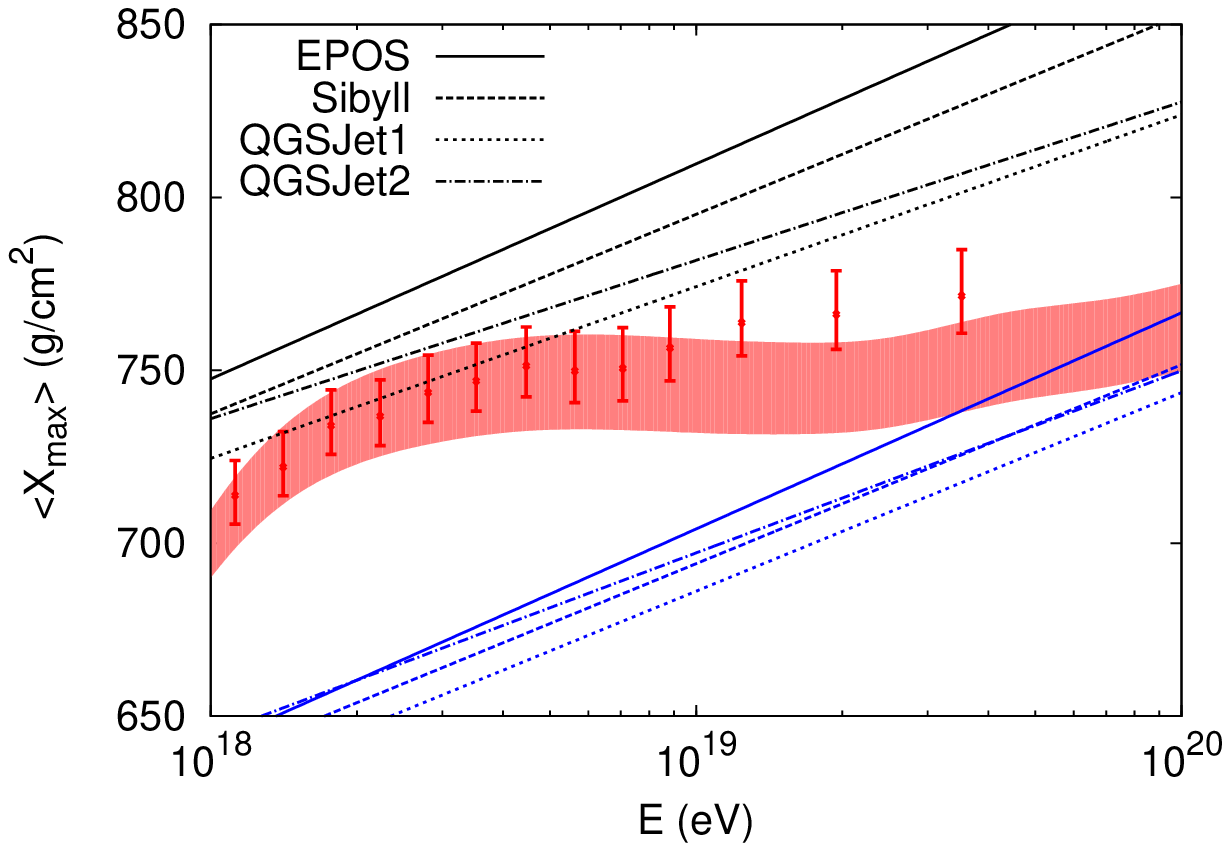}
        \includegraphics[scale=0.46]{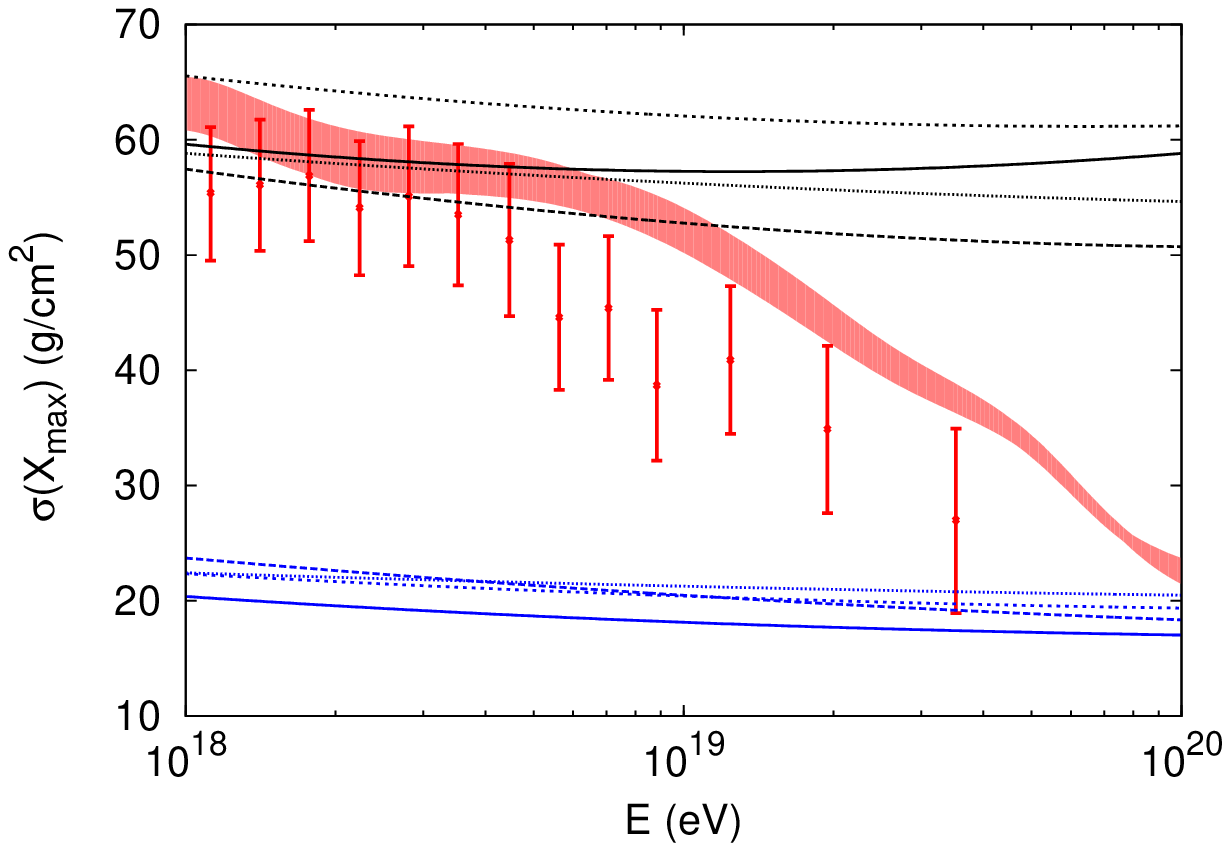}
        \caption{\it Mean value of the depth of shower maximum $\langle X_{max} \rangle$ and its dispersion $\sigma(X_{max})$ as measured by Auger \cite{AugerChem} and in our calculations with the same choice of parameters as in figure \ref{flux}. The shadowed red band corresponds to the uncertainties due to the MC scheme adopted (see text).}
\label{chem}
    \end{center}
\end{figure}

To determine $\langle X_{max} \rangle$ and $\sigma(X_{max})$ we will use the results published in \cite{Abreu:2013env} that provide a simple recipe to compute these two quantities, given a specific choice of the hadronic interaction model used to simulate the shower development in the atmosphere. In \cite{Abreu:2013env} four different MC schemes were considered: EPOS 1.99, Sibyll 2.1, QGSJet 1 and QGSJet 2 \cite{MCcodes}. In the following we will present results on $\langle X_{max} \rangle$ and $\sigma(X_{max})$ as shadowed regions that constrain the results of the four different MC schemes cited, this way of presenting the results is useful to show the uncertainties connected with the hadronic interaction model chosen. Therefore, in our analysis we will use three observables, i.e. flux $J(E)$, $\langle X_{max} \rangle$ and $\sigma(X_{max})$, to constrain theoretical models, i.e. injection power law index $\gamma_g$, source emissivity ${\cal L}_0$ and injection ratios of different nuclei species.  

In figures \ref{flux} (left panel) and \ref{chem} we show our results. As discussed in \cite{Aloisio:2013hya}, in order to reproduce the chemical composition observed by Auger it is needed to assume that heavy elements are injected at the source with a very flat injection spectrum $\gamma_g<1.5$, while light elements (p+He) should be injected with a steep spectrum $\gamma_g>2.5$. This result directly follows from the Auger observation of a chemical composition that is proton dominated at low energies ($<3\div 4\times 10^{18}$ eV) and smoothly drifts toward a heavier composition at higher energies. From the heavy composition at high energies, assuming a rigidity dependent approach $E^Z_{max}=Z E^p_{max}$, follows that the maximum acceleration energy for protons cannot be larger than ${\rm few}\times 10^{19}$ eV, in figure \ref{flux} (left panel) and \ref{chem} we have assumed $E^p_{max}=5\times 10^{18}$ eV in the case of sources providing heavy nuclei and $E^p_{max}=3\times 10^{19}$ eV in the case of sources injecting only light elements. 

This result on the maximum energy surely represents a sort of change of paradigm in the physics of UHECR. While in the past a lot of theoretical efforts were made to model a very high maximum energy ($>10^{20}$ eV), nowadays, after the Auger observations, the situation seems changed with a required maximum energy for protons well below $10^{20}$ eV. 

In figures \ref{flux} (left panel) and \ref{chem} we assumed a source emissivity of the light component, with an injection power law $\gamma_g=2.7$ and composed only by proton and Helium, as ${\cal L}_0(p, He) = 7\times 10^{49}$ erg/Mpc$^3$/yr (above $10^7$ GeV/n) with an injection ratio $Q_{acc}^{He} = 0.1Q_{acc}^p$. The second component, with a flat injection $\gamma_g=1.0$ and contributing p, He, CNO, MgAlSi and Fe, has an emissivity of ${\cal L}_0 = 1.5\times 10^{44}$ erg/Mpc$^3$/yr (above $10^7$ GeV/n) with a ratio of the injected elements as

$$ Q_{He}^{\rm acc}= 0.2 Q_p^{\rm acc},\;  Q_{CNO}^{\rm acc}= 0.06
Q_p^{acc},\; Q_{MgAlSi}^{\rm acc}= 0.03 Q_p^{acc},\; Q_{Fe}^{\rm acc}
= 0.01 Q_p^{acc}. $$

The total fluxes of p and He are plotted as thick continuous green and magenta lines respectively, obtained as the sum of the two contributions to p and He spectra from the two classes of sources considered. At EeV energies sources providing also heavy nuclei give a very small contribution to the flux of p and He, in this energy range the flux of light elements (p+He) is contributed mainly by sources with steep injection. The flux of secondary nuclei, products of the photo-disintegration process, is plotted through the grey shadowed area. For each injected primary specie the total flux (primary plus secondaries) is plotted with continuos colored lines as labeled. In the left panel of figure \ref{flux} the end of the proton spectrum coincides with the maximum energy reached in the sources, while the spectra of nuclei are ended by photo-disintegration on the EBL. Together with the extragalactic CR components, in the left panel of figure \ref{flux} we also plot the tail of the galactic (iron dominated) CR spectrum (black dotted line) as computed in \cite{SelfGen}. 

The hard injection spectra required to fit the Auger data might be reminiscent of models of the origin of UHECRs associated to acceleration in rapidly rotating neutron stars \cite{gamma1}, although hard spectra are a more general characteristic of acceleration scenarios where regular electric fields are available (e.g. unipolar induction and reconnection).

An additional component of extragalactic light nuclei with a generation spectrum much steeper than the one used for heavy nuclei can be introduced making use of the recent data collected by the KASCADE-Grande (KG) collaboration, which show the existence at sub-EeV energies of a light (p+He) component with a spectral index $\gamma_g= 2.79 \pm 0.08$ \cite{KG} attributed to extragalactic sources. Therefore, our hypothesis is compatible with the experimental results of KG, as shown in the right panel of figure \ref{flux} where we plot the KG data points together with the systematic uncertainties (shaded area). The rapidly falling dotted lines in figure \ref{flux} (right panel) show the Galactic (p+He) spectrum as computed in \cite{SelfGen}, with a maximum energy of protons of 5, 6 and 7 PeV (see labels). The roughly constant black dotted line shows the flux of extragalactic light CRs as calculated above, based on the fit to the Auger data. The solid lines indicate the sum of the galactic and extra galactic light components, showing a remarkable agreement with the KG data.

\section{Conclusions}
\label{sec:conclude}

In this paper we took the Auger data on the spectrum and chemical composition of UHECRs at face value and tried to infer as much physical information as possible. The evidence that CRs in the energy region $1\div 5\times 10^{18}$ eV are dominated by light elements may be considered rather solid as it follows from data on $\langle X_{max} \rangle$ and its dispersion for the three largest UHECR detectors, Auger \cite{AugerChem}, TA \cite{TAChem} and HiRes \cite{HiResChem}. Most of the debate on mass composition concentrates upon data at energies $\ge 5 \times 10^{18}$ eV.

Here we showed that the spectrum and chemical composition observed by Auger require hard injection spectra for the heavy component as also discussed in \cite{Andy}. Moreover, we found that the maximum acceleration energy for nuclei of charge $Z$ should be relatively low, namely $\simeq 5Z\times 10^{18}$ eV. From the theoretical point of view the hard injection spectrum is interesting in that it suggests an acceleration mechanism not based on the diffusive shock acceleration paradigm. 

The most disappointing consequence of the hard injection spectra is that the Auger spectrum can only be fitted for energies $\ge 5 \times 10^{18}$ eV, while lower energy CRs imply a different explanation. Filling this gap requires the introduction of an ad hoc CR component and we showed here that such component must be composed of extragalactic light nuclei (p+He) with an injection spectrum with a slope $\gamma_g\simeq 2.7$. The most straightforward implication of this fact is that the transition from galactic to extragalactic CRs must be taking place at energies $\le 10^{18}$ eV rather than at the ankle.

Remarkably this light component has a spectrum and flux which are compatible with the recently detected flux of light nuclei in the energy region $10^{16}\div 10^{18}$ eV by KASCADE-Grande \cite{KG}. These data show an ankle-like feature at $\simeq 10^{17}$ eV, that may be tentatively associated to the transition to extragalactic protons. The disappointing complexity of the viable explanations for the spectrum and chemical composition of Auger are probably the sign that the injection spectra needed to fit the data are themselves the result of a more complex phenomenology. An instance of this could be the propagation in extragalactic magnetic fields and/or phenomena that occur inside the sources that may also potentially affect the spectra of nuclei injected on cosmological scales and possibly preferentially select high energy nuclei.

\section*{Acknowledgements}
The author warmly thanks A.F. Grillo for joint work on the subject of UHECR and his presentation at the Vulcano Conference. The author also thanks V.S. Berezinsky and P. Blasi with whom part of the results presented here were obtained.

\end{document}